\documentclass{article}
\usepackage{spconf,amsmath,graphicx}
\usepackage{algorithm}
\usepackage{algorithmic}
\usepackage{caption}
\usepackage{graphicx}
\usepackage{multicol}
\usepackage{multirow}
\usepackage{float}
\usepackage{subcaption}
\usepackage{mdwlist}
\usepackage{enumitem}
\usepackage{url}

\title{QI-TTS: questioning intonation control For emotional Speech Synthesis}

\name{Haobin Tang$^{1,2 \dagger}$, Xulong Zhang$^{1 \dagger}$, Jianzong Wang$^{1\ast}$, Ning Cheng$^{1}$, Jing Xiao$^{1}$\thanks{$\dagger$ Equal Contribution}\thanks{$^\ast$Corresponding author: Jianzong Wang, jzwang@188.com.}}
\address{$^{1}$Ping An Technology (Shenzhen) Co., Ltd.\\$^{2}$University of Science and Technology of China}
\begin{document}
%
\maketitle

\begin{abstract}
Recent expressive text to speech (TTS) models focus on synthesizing emotional speech, but some fine-grained styles such as intonation are neglected. In this paper, we propose QI-TTS which aims to better transfer and control intonation to further deliver the speaker's questioning intention while transferring emotion from reference speech. We propose a multi-style extractor to extract style embedding from two different levels. While the sentence level represents emotion, the final syllable level represents intonation. For fine-grained intonation control, we use relative attributes to represent intonation intensity at the syllable level. 
Experiments have validated the effectiveness of QI-TTS for improving intonation expressiveness in emotional speech synthesis.  
\end{abstract}


\section{Introduction}
\label{sec:intro}
Due to the rapid advancement of seq2seq modeling architecture, 
style transfer TTS~\cite{skerry2018towards,wang2018style,bian2019multi} has become a prevalent approach for emotional speech synthesis in recent years. The approach utilizes reference audio to specify the desired speech style and its intention is to generate speech that emulates the style of the reference audio.
Reference-based~\cite{skerry2018towards} style transfer involves unsupervised learning of a fixed-length style embedding through expressive samples, which is then utilized to model the speaking style of a reference audio. 
Style transfer has seen significant progress in recent times, with the emergence of numerous approaches such as Global Style Tokens (GST)~\cite{wang2018style}, Variational Autoencoder (VAE)~\cite{liu2020towards,kingma2013auto} and their variants.



However, merely utilizing a learned sentence level style embedding in emotion transfer is insufficient for fully expressing a speaker's attitude. The emotion embedding lacks the ability to model a combination of emotion and mutually exclusive intonation, such as ``angry statement" and ``angry question" accurately at the same time. 
By using questioning intonation, speaker can express a statement in the form of declarative question~\cite{DBLP:conf/interspeech/BaiKZ22}. For example, ``You failed the exam this time." to ``You failed the exam this time?". Thus, it is inadequate to model speech prosody from a single aspect and intonation is essential for intention clarification. There are three issues we would like to address:
1) The existing emotion modeling frameworks only consider the normal statement and lack of the ability to model multiple and differential prosody such as questions in each emotion; 
2) The intonation expressions that exist in human language are nuanced and vary in intensity. Thus we desire to flexibly deliver questioning intonation with specific intensity;
3) limitation of the ability to disentangle prosody from other attributes like content, resulting in the quality degrade and expressiveness instability.

We propose QI-TTS, a model built upon non-autoregressive TTS system FastSpeech2~\cite{ren2020fastspeech} with the following specific designs for above problems. We propose a muti-style extractor that extracts emotion features from sentence level while extracting intonation features from final syllable level to model intonation and emotion at the same time. By utilizing relative attribute~\cite{parikh2011relative}, we are able to assign a relative degree of intonation strength to each audio samples in dataset via a ranking function.
To minimize the information overlap between content embedding and style embedding, we utilized a gradient reversal layer (GRL) together with content predictor.

The paper presents the following contributions:
1) We jointly transfer the emotion and intonation from reference audio in an end-to-end way which further delivers the speaker's intention;
2) QI-TTS is capable of learning the variation of intonation intensity in speech without the use of explicit labels, making it possible to control the intonation intensity effectively using either manual instructions or reference speech;
3) Experiments have validated the effectiveness of QI-TTS for expressive style transfer and intonation intensity control. 

\begin{figure*}[htbp]   
\centering
\includegraphics[width=0.7\linewidth]{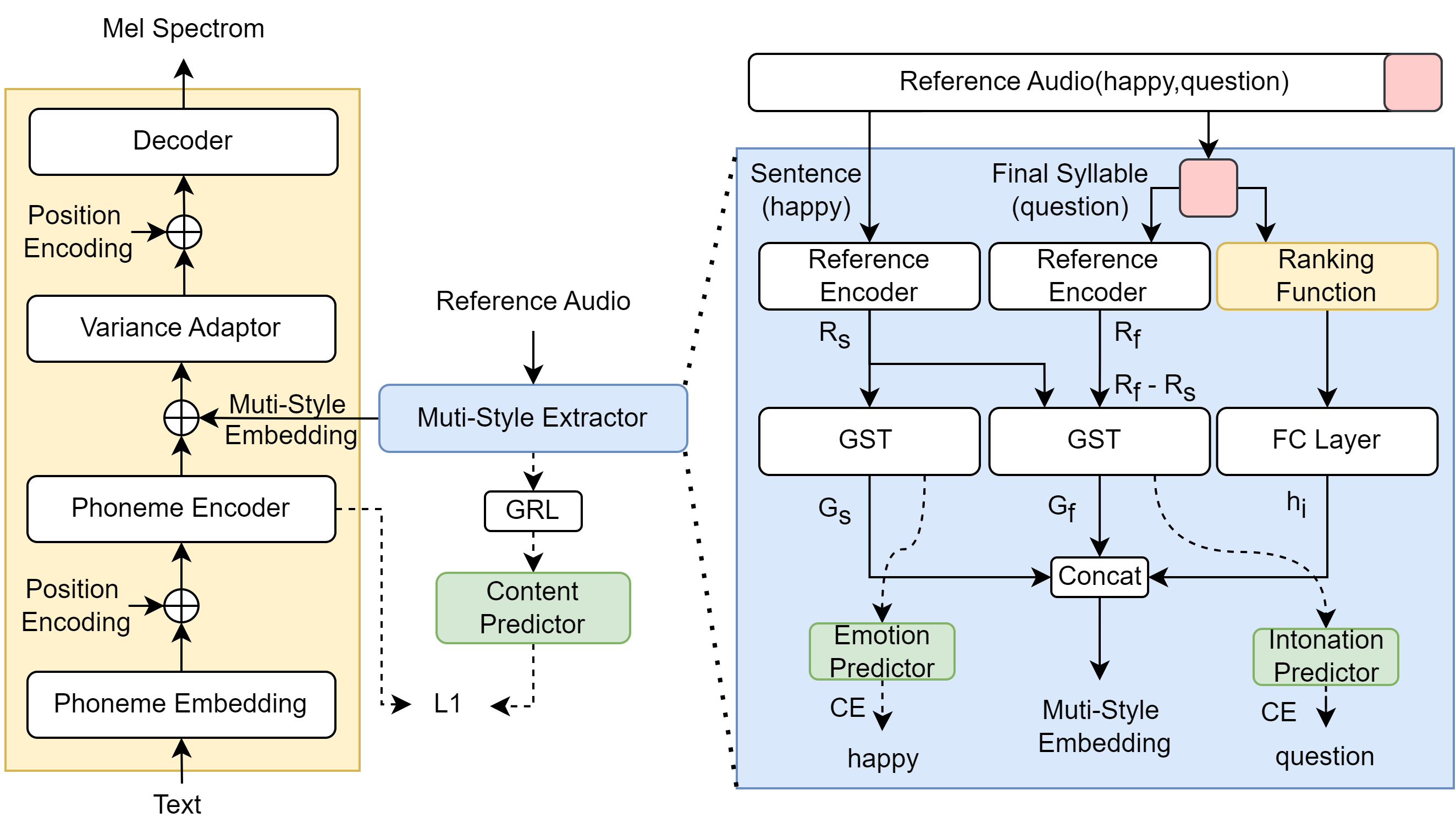}
\caption{The overview architecture for QI-TTS. The red part of reference audio refers to final syllable. "GRL" denotes gradient reversal layer. $R_s$ and $R_f$ denote the reference embedding of the reference sentence and that of the final syllable, respectively. $G_s$ and $G_f$ denote emotion and intonation embedding. $h_i$ represents intonation intensity embedding.}
\label{Fig1}
\end{figure*}

\section{Proposed Method}
\label{sec:method}
As Fig.~\ref{Fig1} depicts, the overall architecture of QI-TTS can be mainly divided into two parts based on FastSpeech 2~\cite{ren2020fastspeech}, a muti-style extractor with ranking function, and a content predictor with gradient reversal layer (GRL)~\cite{ganin2016domain}. Specifically, the extractor is used to extract emotion and intonation embedding at sentence and final syllable level. The ranking function outputs a relative intonation intensity and encodes it into an intonation intensity embedding which is concatenated with emotion and intonation embedding to form muti-style embedding. The GRL content predictor is used to further disentangle content information from muti-style embedding.    

\subsection{Muti-Style Extractor}
Linguists find that the duration and fundamental frequency in sentence-final differ from that in other positions of the utterance~\cite{berkovits1984duration}. The duration of final syllable is 1.53 times longer than that of non-final syllable in English~\cite{delattre1966comparison}. The final syllable's f0 valley is a significant feature of a statement. Furthermore, the absolute value of the endpoint F0 was the strongest cue in distinguishing statements from questions rather than the slope of the terminal glide~\cite{majewski1969influence}. 
Research~\cite{berkovits1984duration} shows that the mean of final syllable in English is 0.37 seconds with a standard deviation of 0.15. So we model the last 0.52 seconds of the audio that contains the final syllable as intonation to capture the duration variance and intonation related features.

We designed a multi-style extractor, depicted in the blue section of Fig.~\ref{Fig1}, to extract the emotion and intonation information from the reference speech.
The module contains reference encoders and style token layers that follow the structure proposed by~\cite{wang2018style}. Both the Mel-spectrogram of the sentence and the Mel-spectrogram of its final syllable are fed into their corresponding reference encoder. $R_s$ and $R_f$ denote the output sentence reference embedding and final syllable reference embedding repectively.
However, the effect of final syllable reference embedding overlaps with that of sentence reference embedding, since the relationship between emotion and intonation is not completely hierarchical. To reduce such overlapping, we use $R_f - R_s$ to represent residual embedding of final syllable level.
The emotion style embedding $G_s$ and intonation style embedding $G_f$ are formed by passing $R_f$ and $R_f - R_s$ to the corresponding style token layers. These embeddings are then concatenated with a ranking embedding that represents intonation intensity to create a multi-style embedding.


\subsection{Modelling Intonation Intensity}
We aim to conduct fine grained control of questioning intonation but the intensity labels are not readily available. A ranking-based method called relative attributes~\cite{parikh2011relative} is used for unsupervised intensity modelling. 
We regard the questioning intensity as a speech attribute which can be depicted by learned relative attributes.
The questioning intensity of a sentence should be zero since it lacks any questioning intonation variation. Therefore, we consider questioning intensity to be a relative difference between statement and question.   
Assuming we have a training set $T$ = ${X_t}$, and $X_t$ denotes the $t^{th}$ training sample's acoustic features. $A$ and $B$ are the question and statement set respectively. We aim to learn the following ranking function:
\begin{equation}
f(X_t) = W{X_t}
\end{equation} 
where $W$ is a weighting matrix that we need to learn. Considering the intonation intensity of question should always higher than that of statement we have to satisfy the following constraints:
\begin{align}
&\forall (X_a\in{A} \ and \ {X_b} \in{B})  : W{X_a} > W{X_b} \nonumber\\
&\forall(X_a,X_b)\in{A} \  or \  (X_a,X_b) \in{B} : W{X_a} = W{X_b}
\end{align}
To estimate the weighting matrix $W$, we solve the following problem~\cite{chapelle2007training}:
\begin{align}
&\mathop{min}\limits_{W}(\frac{1}{2}||W||^2_2 + C(\sum{\xi^2_{a,b}} + \sum{\gamma^2_{a,b}})) \nonumber \\
s.t. &W(X_a - X_b) \geq 1 - \xi_{a,b} ; \forall (X_a\in{A} \ and \ {X_b} \in{B}) \nonumber\\
&|W(X_a - X_b)|\leq\gamma_{a,b};\forall(X_a,X_b)\in{A} \  or \  (X_a,X_b) \in{B} \nonumber\\
&\xi_{a,b}\geq0;\gamma_{a,b}\geq0
\end{align} 
where $C$ is the trade-off between the margin and the size of slack variables $\xi^2_{a,b}$ and $\gamma^2_{a,b}$. Once the relative ranking function
$f(x)$ is trained, a normalized relative attribute in range [0, 1] can be calculated for a speech sample.
This attribute is subsequently fed to a FC layer, which yields intensity embedding $h_i$.
Fine-grained intonation control can be achieved during the inference stage through the use of reference speech or manual instructions. Specifically, the intensity can be predicted by analyzing the reference audio or assigned a value manually within the interval of [0, 1].


\subsection{Prediction Tasks}
The muti-style extractor unsupervisedly learns multi-level styles from reference audio. We add emotion and intonation prediction tasks to force each level module to pay more attention to learning the corresponding style. The cross-entropy (CE) loss is used for emotion predictor. We use the weighted cross entropy function as intonation loss function because of the sparse question labels: 
\begin{equation}
L = -\hat{y_1}logy_1 - \sigma \hat{y_2}logy_2
\end{equation} 
where [$\hat{y_1}$, $\hat{y_2}$] is the probability of the ``statement" and ``question" categories respectively. 
One-hot encoding of ground-truth label is represented by [$y_1$, $y_2$].
We adjust $\sigma$ to ensure the data balance in the training.

An adversarial content predictor network inspired by Mask-And-Predict (MAP)~\cite{wang2021adversarially} is designed to disentangle overlapped content information in muti-style embedding. The network is composed of a gradient reverse layer and a content predictor~\cite{liu2020mockingjay}. The predictor is trained to predict content representation as accurate as possible by minimizing the loss: $L_{content} = {||\hat{c}-c||}_1$, where
the content embedding generated by the phoneme encoder and the output of the adversarial content predictor are denoted as $c$ and $\hat{c}$, respectively. The gradient is reversed before backward propagated to the muti-style extractor to minimize the content information contained in the multi-style embedding. 

\section{Experiment}
\label{sec:Experiment}
\subsection{Training Setting}
To train our model, we use the english part of ESD dataset~\cite{zhou2021seen}, which consists of five emotions spoken by 10 native English (5 male and 5 female): \emph{Neutral}, \emph{Sad}, \emph{Happy}, \emph{Angry}, and \emph{Surprise}. We adopt the same data partition as provided in ESD. Importantly, we add "statement" and "question" labels with the help of K-means method similar to Into-TTS~\cite{lee2022into} and there are 310 questions for each speaker on average.

We follow an a public version of relative attribute~\cite{parikh2011relative} to train the ranking function for questioning intonation intensity. In practice, final syllable's acoustic features of statements and questions are used for calculating intensity. The acoustic features are extracted by the openSMILE~\cite{eyben2010opensmile}. 
We train QI-TTS for 350k iterations with 16 batch size. 
In our experiments, we employ Hifi-Gan~\cite{kong2020hifi} as the vocoder for synthesizing waveforms from the generated Mel-spectrograms.

\begin{table*}[ht]
    \vspace{-1.5em}
	\centering
	\caption{Subjective and objective evaluation results.}
	\scalebox{0.85} {
		\begin{tabular}{lcccccc}
			\hline\hline
			\textbf{Model} & \textbf{MOS $\uparrow$}  & \textbf{SMOS $\uparrow$} & \textbf{Intonation $\uparrow$} & \textbf{MCD $\downarrow$} & \textbf{FFE    $\downarrow$} & \textbf{ Duration MSE $\downarrow$} \\
			\hline 
			 GT & 4.47 ± 0.08 & $/$ & $/$ & $/$ & $/$ & $/$\\
			 GTmel + Vocoder & 4.40 ± 0.09 &  4.47 ± 0.10 & 99.2$\%$ &  2.40 & 0.07 & 0.031\\
			\hline
			 MutiEmo FS2 ~\cite{cui2021emovie}& 3.81 ± 0.08 & 3.85 ± 0.08 & 81.6$\%$  & \textbf{3.15} &  0.43 & 0.144 \\
			 Styler~\cite{lee2021styler} & 3.76 ± 0.08 & 3.97 ± 0.08 & 85.9$\%$  & 5.57 &  0.41 & 0.149  \\
			\hline
			 QI-TTS & \textbf{3.84 ± 0.10} & \textbf{4.01 ± 0.08} & \textbf{95.2}$\%$  & 4.89 &  \textbf{0.39} & \textbf{0.141}\\
			\hline\hline
		\end{tabular}
	}
	\label{tab1}
	\vspace{-1em}
\end{table*}

\subsection{Result Evaluation}
Our comparison involves the audio samples generated by QI-TTS and the systems listed below: 
1) GT, Mel-spectrogram of reference audio; 
2) GT mel + Vocoder, speech samples generated by HiFi-GAN using ground truth Mel-spectrogram; 
3) MultiEmo FS2~\cite{cui2021emovie}, which adds the emotion d-vectors to Fastspeech 2. "FS2" denotes Fastspeech 2; 
4) Styler~\cite{lee2021styler}, a speech synthesis model that employs speech decomposition to represent the style and achieve expressiveness.

For subject evaluation, we assess the quality and expressiveness of the synthesized audios via mean opinion score (MOS) and similarity mean opinion score (SMOS) tests. Subjects are asked to rate 5 question and 5 statement speech in each emotion on a scale from 1 to 5 with 1 point intervals. Besides, the accuracy of intonation perception is taken to evaluate the intonation transfer. Participants are tasked with assessing whether the audio sample is a question or a statement. We calculate the objective matrixes (\textit{i.e.}, mel cepstral distortion (MCD)~\cite{kubichek1993mel}, F0 Frame Error (FFE), and Duration MSE) for objective evaluation. 
MCD evaluates the spectrum similarity while FFE is the proportion of frames with error F0. MSE between the ground-truth duration and predicted one is computed for duration. As depicts in Table~\ref{tab1}, 
QI-TTS is capable of producing speech samples that more closely resemble the intonation of the reference audio with no hinder to emotion transfer, clearly reflecting the correct pitch and duration.


\begin{figure}[!b]
	\centering
	\begin{subfigure}{0.3\linewidth}
		\centering
		\includegraphics[width=\linewidth]{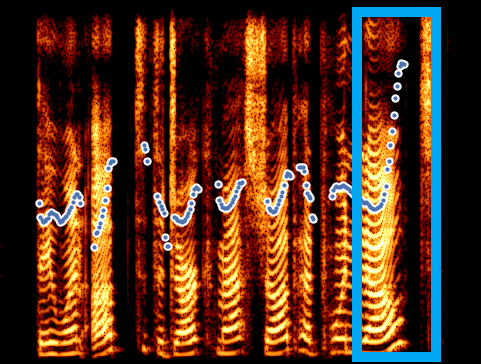}
		\caption{Ground Truth}\label{fig2:a}		
	\end{subfigure}
	\hspace{0.2cm}
	\begin{subfigure}{0.3\linewidth}
		\centering
		\includegraphics[width=\linewidth]{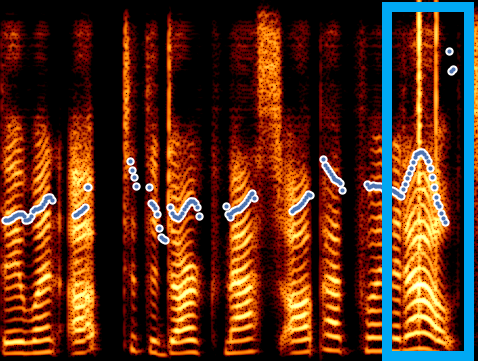}
		\caption{w/o final syllable}\label{fig2:b}
	\end{subfigure}
	\hspace{0.2cm}
	\begin{subfigure}{0.3\linewidth}
		\centering
		\includegraphics[width=\linewidth]{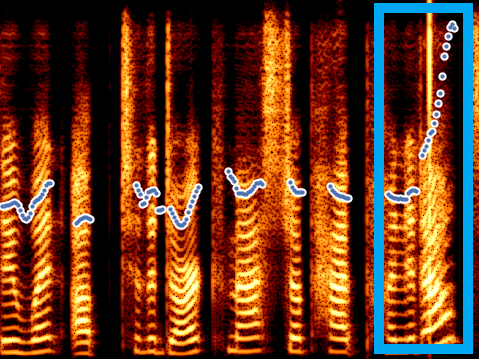}
		\caption{QI-TTS}\label{fig2:c}
	\end{subfigure}
         \caption{Visualizations of a declarative question's Mel-spectrograms in ablation study.}
 \label{fig2:}
\end{figure}
\subsection{Ablation Study}
The effectiveness of techniques employed in QI-TTS, such as final syllable level style, residual style embedding, and predictors, were demonstrated through ablation studies.
To compare the expressiveness of the synthesized speeches, CSMOS (comparative similarity mean opinion score) results are presented in Table~\ref{table2}. Fig.~\ref{fig2:} shows that the model without final syllable level performs wrong intonation while the QI-TTS performs well indicating the importance of modeling intonation style representation in final syllable level. The absence of intonation related function results in more severe decline in question than in statement which demonstrates the efficiency in modeling questioning intonation. Moreover, removing the residual method results in -0.08 CSMOS suggesting that minimizing interference and overlap between emotion and intonation styles is neccesary. The degradation of audio similarity also demonstrates the efficacy of the predictors in accurately modeling specific emotions and intonations while reducing the impact of textual content. 
\begin{table}[ht]
    \vspace{-0.5em}
	\centering
	\caption{CSMOS comparison for ablation study.}
	\scalebox{0.85} {
		\begin{tabular}{p{4.2cm}|p{1.3cm}|p{1.3cm}}
			\hline\hline
			\textbf{Model} & \textbf{Question}& \textbf{Statement} \\
			\hline 
			\quad QI-TTS & / & / \\
			\hline
			\quad w/o final syllable level & -0.15 & -0.09  \\
			\quad w/o residual style & -0.08 & -0.08  \\
			\quad w/o Emotion predictor & -0.10 & -0.10  \\
			\quad w/o Intonation predictor & -0.11 & -0.04 \\
			\quad w/o GRL content predictor & -0.08 & -0.09 \\
			\hline\hline
		\end{tabular}
	}
	\label{table2}
	\vspace{-2em}
\end{table}

\subsection{Intonation Intensity Control}
To evaluate the intonation intensity control, three distinct intensity values were chosen: 0.3, 0.6, and 0.9 which correspond to weak, medium, and strong intensity levels, respectively. 
Best-worst scaling (BWS) test is conduct to evaluate intonation and emotion perception of generated audio with different questioning intensities in each emotion category. The assessors are requested to select the speech sample that best and worst represents a specific emotion or intonation based on their perception. There is little difference between the experimental results of angry, happy, and sad, so we show angry and surprise only in Table~\ref{table3}.

We first evaluate the perception of the questioning intonation. 
According to Table~\ref{table3:a}, the highest ``Best" score for question intonation always occurs at 90$\%$ questioning intensity, and this score increases with increasing intensity.
Conversely, the highest ``Worst" score is always observed at the lowest percentage of questioning intonation. These findings provide empirical support for the effectiveness of controlling intonation intensity.
We further evaluate emotion perception in synthesized speech to study the effect of intonation intensity control on emotional expression in Table~\ref{table3:b}.
The “Best” score of “Surprise” slightly increases as the percentage of questioning intonation intensity increases while that of “Angry” slightly decreases. We attribute this phenomenon to the potential link between surprises and questions. Stronger question helps express surprise, but excessive transfer confuses other emotions with surprise to some extent. 
\begin{table}[htbp]
    \caption{Best-worst scaling (BWS) test for questioning intonation and emotion perception.} 
    \label{table3}
    \begin{subtable}{\linewidth}
      \centering
        \caption{Perception of questioning intonation}
        \label{table3:a}
        \scalebox{0.85}{
    		\begin{tabular}{p{2.2cm}|p{2.2cm}|p{1.2cm}|p{1.2cm}}
    			\hline\hline
    		    \multicolumn{2}{c|}{\textbf{Configuration}}& \textbf{Best$(\%)$} & \textbf{Worst$(\%)$}\\
    		    \hline
    			\quad \multirow{3}{*}{Surprise} & 30$\%$ Question & 8 & 79 \\
    			\quad & 60$\%$ Question & 11 & 21\\
    			\quad & 90$\%$ Question & 81 & 0 \\
    			\hline\hline
    			\quad \multirow{3}{*}{Angry} & 30$\%$ Question & 8 & 69 \\
    			\quad & 60$\%$ Question & 15 & 31\\
    			\quad & 90$\%$ Question & 77 & 0 \\
    			\hline\hline
    		\end{tabular}
        }
    \end{subtable} 
    
    \begin{subtable}{\linewidth}
      \centering
        \caption{Perception of emotion }
        \label{table3:b}
        \scalebox{0.85}{
    		\begin{tabular}{p{2.2cm}|p{2.2cm}|p{1.2cm}|p{1.2cm}}
    			\hline\hline
    		    \multicolumn{2}{c|}{\textbf{Configuration}}& \textbf{Best$(\%)$} & \textbf{Worst$(\%)$}\\
    		    \hline
    			\quad \multirow{3}{*}{Surprise} & 30$\%$ Question & 29 & 39 \\
    			\quad & 60$\%$ Question & 34 & 33\\
    			\quad & 90$\%$ Question & 37 & 28 \\
    			\hline\hline
    			\quad \multirow{3}{*}{Angry} & 30$\%$ Question & 39 & 28 \\
    			\quad & 60$\%$ Question & 40 & 27\\
    			\quad & 90$\%$ Question & 21 & 45 \\
    			\hline\hline
    		\end{tabular}
        }
    \end{subtable}%
\end{table}

\section{Conclusion}
This paper proposes QI-TTS, a multi-style transfer model, which can better transfer emotion and intonation from reference audio and achieve intonation intensity control for expressive TTS.  Experimental results demonstrate that QI-TTS performs better on expressive speech synthesis and intonation intensity control missions. In future research, we will explore the effectiveness of QI-TTS in multilingual scenarios.

\section{Acknowledgement}
Supported by the Key Research and Development Program of Guangdong Province (grant No. 2021B0101400003) and corresponding author is Jianzong Wang (jzwang@188.com).


\vfill\pagebreak
\bibliographystyle{IEEEbib}
\bibliography{QITTS}

\end{document}